# A Tutorial for Analyzing Structural Equation Modelling

Hashem Salarzadeh Jenatabadi
Department of Science and Technology Studies
University of Malaya, Kuala Lumpur, Malaysia
salarzadeh@um.edu.my
+60178777662

**Abstract**

This paper provides a tutorial discussion on analyzing structural equation modelling (SEM). SEM can be regarded as regression models with observed and unobserved indicators, have been extensively applied to practical and fundamental studies. We deliver an introduction to SEM method and a detailed description of how to deal with analyzing the data with this kind of modelling. The intended audience is statisticians/mathematicians/methodologists who either know about SEM or simple basic statistics especially in regression and linear/nonlinear modeling, and Ph.D. students in statistics, mathematics, management, psychology, and even computer science.

**Keywords:** Structural Equation Modeling, General Linear Model, Regression

**Introduction**

SEM is a strong statistical methodology that combines statistical data and qualitative causal assumption to assess and evaluate the causal associations. SEM can effectively replace multiple regression, covariance analysis, time series analysis, factor analysis, and path analysis, which means that it is possible to consider the said procedures as special cases of SEM. In other words, SEM is considered as an extension of the general linear model (GLM) of which multiple regression modelling is a part.

The current study is based on a special case of SEM, which was used as path analysis through which the research hypotheses were examined and evaluated. The application and role of the SEM technique, i.e., path analysis, are explained in detail in the following section. Furthermore, in the following section the two-steps process, which is one of the most commonly used, is elaborated upon. However, the limitations and shortcomings of the said process will be determined and analysed, and, in the end, path analysis will be suggested as the recommended and commonly used technique employed for analysis of the data.

To describe SEM schematically, it can be portrayed as a model that uses particular configurations of the structures of four graphical symbols, that is, an ellipse (or circle), a rectangle, and a single or "double-headed arrow". Generally, squares (or rectangles; 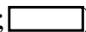) and circles (or ellipses; 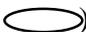) show observed and unobserved (latent) variables respectively, "single-headed arrows" (→) represent the direction of the impact of one factor on another, and "double-headed arrows" (↔) display correlations or covariance that take place between the variable or indicator pairs. The four symbols mentioned are utilized by researchers within the four basic configuration frameworks to create a specific structural model. Each of the four basic configurations is a vital component in the analysis process. A brief description of each of the four configurations is presented below:

- 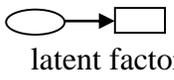 Shows an observed factor's regression path coefficient onto an unobserved or latent factor.
- 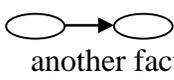 Represents a regression model's path coefficient of one factor or variable onto another factor or variable.
- 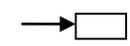 Displays the term of measurement error connected to an observed factor.
- 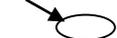 Represents the residual error term in an unobserved or latent variable's prediction.

**Key Concepts in SEM**

This section defines some essential terms and concepts that have been used or referred to in this study. For the presented definitions the researcher has used a variety of resources, which have been included in the list of references used in this thesis.

Indicators (manifest or reference variable) are an observed variable like items used in a survey. Unobserved (latent) variable/factor/construct is an unobserved (latent) variable that is measurable by its respective indicator. The unobserved variable can be a dependant variable or mediator. An exogenous variable is a totally independent factor without any prior causal factor, which might be associated with the exogenous variable. This association is graphically displayed with a double-headed arrow. The endogenous variable is a completely dependent factor and mediator, which is both the effect of other exogenous factors and cause of other endogenous factors. The mediator is both the effect of other exogenous factors and cause of other dependent factors.

The variables of a model, based on the role of their effects and causes can be categorized into upstream and downstream variables, respectively. The unobserved variables representation depends on their relationship with the observed indicator variables, which is regarded to be one of the essential characteristics of SEM.

The following diagram shows the observed variables or indicators with M1 to M9, and the latent variables with Q1 to Q3:

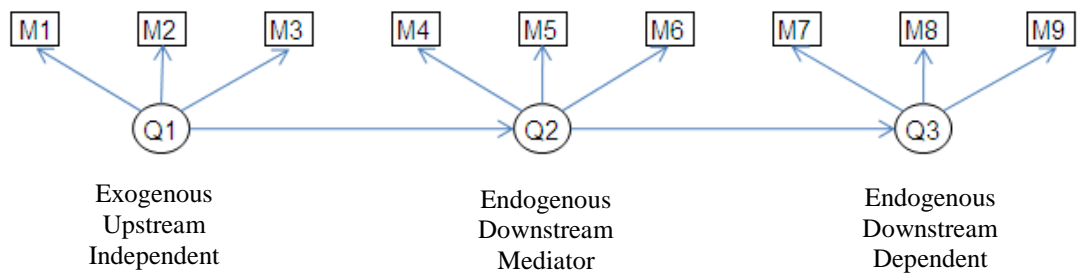

**Why SEM Method?**

As introduced and discussed in Chapter One of this thesis, the hypotheses were tested and checked through path analyses. Path analysis is a particular type of SEM that is itself a GLM development. GLM is a second generation of the method of data analysis, which depends on a structural relationship existing among variables of interest. SEM can be carried out using software – AMOS (Analysis of Momentum Structures), LISREL, Mplus, EQS – that are available and accessible to the researcher. These software packages were employed to evaluate and assess the relationship among the collected data, manifest, i.e., conceptual model including observed variables and latent hypothetical factors, i.e., latent constructs or unobserved variables [1].

Panel data, regression, and analysis of variance (ANOVA) are the most methodologies that have been using in so many studies [2-7]. In recent years, SEM has attracted the attention of many researchers and organizations as a commonly adopted method used for tasks like data analysis in various disciplines like airline [8-12], education [13, 14], management [15-20], and computer science [21, 22]. Nevertheless, the popularity of SEM relies on some of the advantages that form part of the features of this method. Some of these properties are reviewed below:

The first and the most significant of SEM advantages is its ability to enable the researcher to simultaneously model and examine the indirect and direct interrelationships that exist among multiple dependent and independent [23]. This feature is a vital ability in the current research in which the model has an essential factor of mediation, i.e., a dependent variable, such as internal operation or performance, which, in the subsequent independent relationships, changes into an independent variable. After the effects of all other paths are taken into consideration, SEM assesses and evaluates each path coefficient separately. This feature makes SEM the most appropriate means for testing and estimating the role of the variables. As a result, the impact of a predictor factor is conveyed to a standard playing the role of a mediator. In fact, SEM is an effective and optimal technique for checking and testing the relationships among mediator variables [24, 25].

Another characteristic that makes SEM the preferred model compared to methods of conventional multiple regressions is its typically piecemeal nature in generating separate and individually distinct coefficients. The SEM technique permits the researcher to check and examine a complete model generating goodness-of-fit statistics and assessing the overall fit of the complete model [26]. The next feature of SEM allows the expansion of statistical estimating by the researcher through assessing and estimating terms of error for observed variables. In the traditionally and conventionally employed multivariate processes, such as multiple regression modelling, the error rate of variables measurement and the between variables residuals or their observed variables, i.e., indicators, are null [27, 28]. However, this sort of assumption does not look realistic because the gauged variables usually have some measurement errors, even if small. Consequently, biased coefficients are expected to result from the utilization of these kinds of measurements, which is usual in conventional multivariate methods. Nevertheless, SEM enables the researcher to apply terms of measurement error to the process of estimation, which, ultimately, contributes to the improvement of the structural path coefficients reliabilities [29].

Yet another feature of SEM, which distinguishes it from other available models is its ability to allow the researcher to incorporate both observed (manifest) and unobserved (latent) variables into the process of the same analysis. As a result, the created incorporation provides a stronger analysis of the suggested model as well as a better evaluation of the study [23, 29]. Furthermore, SEM has the ability to assist researchers in two more ways, i.e., handling complicated data (with non-normality and multicollinearity) and use of modelling of graphical interfaces [30].

Concisely, the important characteristics that make SEM more preferred in comparison with other available conservative multivariate methods, such as multiple regression modelling, is its ability to allow researchers to model the mediator variables to check and test the models with multiple dependent and independent indicators; to model mediator factors, and to analyse whole systems of indicators that enable the researcher to simultaneously establish models with a more realistic nature that need simultaneous analysis[31, 32].

**Procedure of SEM**

SEM has a two-step procedure, first of which concerns the measurement model validating and the second step is about the assumed structural model testing [33]. The first step, measurement model, deals with the relationships between unobservable (latent) and observable (measurement) factors [30]. In other words, the measurement model concerns one part or all parts of an SEM relating to the unobservable variables and their indicators. However, confirmatory factor analysis [34] is utilized for primary operation of the measurement model. Therefore, it can be easily said that the CFA model is a pure measurement model containing un-gauged covariance between each of the possible latent variable pairs.

The outcome from this procedure is goodness of fit values applicable to further enhance the measurement scales level, that is, indicator variables, through gauging the related latent constructs [35]. If the measurement model's goodness of fit measures are satisfactory, i.e., where the measurement model can provide the required data with a goodness of fit, then it can be concluded that the indicators' targeted constructs can be measured adequately. However, if the measurement model is not able to provide a sufficiently powerful fit to the data, it can be then concluded that, at least some of the observed factors are unreliable. In this situation, prior to structural model analysis, it is required to refine the scales of the measurement anew [33]. Otherwise, moving to the structural model will not be of any use unless the model is confirmed as a valid model with satisfactory results.

The structural model also specifies the structural relationships, that is, indirect and direct impacts among unobservable constructs. As stated earlier in this chapter, for testing the structural model, some statistical software and packages, such as AMOS, LISREL, Mplus, and EQS are applicable. Nevertheless, the structural model, also known as the default model, can be contrasted against the measurement model.

A set of endogenous and exogenous factors, the direct effects (arrows with single-head) linking these variables together, and the error terms for these factors (reflecting the unmeasured factors effects that are not included in the research model) are contained in this model. The statistics of goodness of fit is another outcome of this analysis. The goodness of fit statistics can be employed to evaluate and judge the whole model and the hypotheses, as well as measure how much the expected covariances can be adjusted to the observed covariances in the data. In addition, other productions of this analysis are Modification Indexes (MIs) that can be applied to improvement of the model fit, i.e., to fit the model to the observed research data. However, application of MIs is based on hypothetical considerations [30].

In summary, the SEM technique is normally carried out in two phases, namely: 1) validating the factors of the latent indicators construct, i.e., the scale of measurement, (CFA evaluated the measurement model; and 2) the structural model procedure is evaluated to judge the whole fitting model as well as the individual structural models hypothesized among the unobservable (latent) indicators.

**SEM Software Packages**

Various software packages are available that are compatible with SEM that can be executed on home or office PCs. Some of these programs are the CALIS procedure of SAS/STAT, AMOS, EQS, MPLUS, LISREL, RAMONA module of SYSTAT, MX GRAPH, and the SEPATH module of STATISTICA. The main difference among them is in their ability and capability to back more sophisticated analyses and interaction methods with the program. The specifications and capabilities of such programs, like any other computer program, are likely to alter with the release of their newer versions; hence, the computer programs cannot be easily described save for the analysis outcomes of the model and a short explanation of the output and its description. In the current study, AMOS and SPSS are the selected programs for the statistical analysis of performance data.

**Discussion**

The main structure SEM is similar to multiple regression, and includes multiple dependent and independent variables, however, SEM acts in a stronger and more effective way by taking into consideration the modelling of correlated error terms, interactions, nonlinearities, correlated independents, measurement error, one or more latent dependent variables with multiple indicators, and multiple latent independent variables, which are measured by multiple indicators. The SEM method can be utilized as a stronger and more effective option for path analysis, factor analysis, multiple regression, analysis of covariance, and time series method [36]. This means that these procedures are possible for consideration as specific cases of SEM, in other words, SEM can be considered as an extension of the earlier model known as generalized linear modelling (GLM), which has multiple regression as a part of itself.

**References**
1. Hoyle, R.H. and G.T. Smith, *Formulating clinical research hypotheses as structural equation models: A conceptual overview.* Journal of Consulting and Clinical Psychology, 1994. **62**(3): p. 429-440.
2. Jenatabadi, H.S. and N.A. Ismail, *The Determination of Load Factors in the Airline Industry.* International Review of Business Research Papers, 2007. **3**(4): p. 125-133.
3. Jenatabadi, H.S. and A. Noudoostbeni, *End-User Satisfaction in ERP System: Application of Logit Modeling.* Applied Mathematical Sciences, 2014. **8**(24): p. 1187-1192.

4. Samimi, P. and H.S. Jenatabadi, *Globalization and Economic Growth: Empirical Evidence on the Role of Complementarities.* PloS one, 2014.
5. Noudoostbeni, A., N.M. Yasin, and H.S. Jenatabadi. *To investigate the success and failure factors of ERP implementation within Malaysian small and medium enterprises.* in *Information Management and Engineering, 2009. ICIME'09. International Conference on.* 2009. IEEE.
6. Noudoostbeni, A., N.M. Yasin, and H.S. Jenatabadi. *A mixed method for training ERP systems based on knowledge sharing in Malaysian Small and Medium Enterprise (SMEs).* in *Information Management and Engineering, 2009. ICIME'09. International Conference on.* 2009. IEEE.
7. Noudoostbeni, A., et al., *An effective end-user knowledge concern training method in enterprise resource planning (ERP) based on critical factors (CFs) in Malaysian SMEs.* International Journal of Business and Management, 2010. **5**(7): p. P63.
8. Ismail, N.A. and H.S. Jenatabadi, *The Influence of Firm Age on the Relationships of Airline Performance, Economic Situation and Internal Operation.* Transportation Research Part A: Policy and Practice, 2014.
9. Jenatabadi, H.S., *Impact of Economic Performance on Organizational Capacity and Capability: A Case Study in Airline Industry.* International Journal of Business and Management, 2013. **8**(17): p. p112.
10. Jenatabadi, H.S., *Introduction Latent Variables for Estimating Airline Assessment.* International Journal of Business and Management, 2013. **8**(18): p. p78.
11. Jenatabadi, H.S. and N.A. Ismail, *A NEW PERSPECTIVE ON MODELING OF AIRLINE PERFORMANCE.* 3rd International Conference on Business and Economic Research, March 12-13 2012, Bandung, Indonesia., 2012.
12. Jenatabadi, H.S. and N.A. Ismail, *Application of structural equation modelling for estimating airline performance.* Journal of Air Transport Management, 2014. **40**: p. 25-33.
13. Hui, H., et al., *Principal's leadership style and teacher job satisfaction: A case study in China.* 2013.
14. Dadkhah, V., H. Hui, and H.S. Jenatabadi, *An Application of Moderation Analysis: the Situation of School Size in the Relationship among Principal's Leadership Style, Decision Making Style, and Teacher Job Satisfaction.* International Journal of Research in Business and Technology, 2014. **5**(3): p. 724-729.
15. Hui, H., H.S. Jenatabadi, and S. Radu. *Knowledge Management and Organizational Learning in Food Manufacturing Industry.* in *International Conference on Economic, Finance and Management Outlooks (ICEFMO 2013). October 5-6 2013, Kuala Lumpur, Malaysia.* 2013. Kuala Lumpur, Malaysia.
16. Hui, H., et al., *Influence of Organizational Learning and Innovation on Organizational Performance in Asian Manufacturing Food Industry.* Asian Journal of Empirical Research, 2013. **3**(8): p. 962-971.
17. Hui, H., et al., *The Impact of Firm Age and Size on the Relationship Among Organizational Innovation, Learning, and Performance: A Moderation Analysis in Asian Food Manufacturing Companies.* Interdisciplinary Journal of Contemporary Research In Business, 2013. **5**(3).
18. Jasimah, C.W., et al., *Explore Linkage between Knowledge Management and Organizational Performance in Asian Food Manufacturing Industry.* International Journal of Asian Social Science, 2013. **3**(8): p. 1753-1769.
19. Jenatabadi, H.S., *Situation of Innovation in the Linkage between Culture and Performance: A Mediation Analysis of Asian Food Production Industry.* Contemporary Engineering Sciences, 2014. **7**(7): p. 323-331.
20. Radzi, C.W.J.W.M., et al., *The Relationship among Transformational Leadership, Organizational Learning, and Organizational Innovation: A Case Study in Asian Manufacturing Food Industry.* Asian Journal of Empirical Research, 2013. **3**(8): p. 1051-1060.

21. Jenatabadi, H.S., *An Application of Moderation Analysis in Structural Equation Modeling: A Comparison Study between MIS and ERP.* Applied Mathematical Sciences, 2014. **8**(37): p. 1829 - 1835.
22. Jenatabadi, H.S., et al., *Impact of supply chain management on the relationship between enterprise resource planning system and organizational performance.* International Journal of Business and Management, 2013. **8**(19): p. p107.
23. Gefen, D., D. Straub, and M.C. Boudreau, *Structural equation modeling and regression: Guidelines for research practice.* Communications of the Association for Information Systems, 2000. **4**(1): p. 7.
24. Dhanaraj, C., et al., *Managing tacit and explicit knowledge transfer in IJVs: the role of relational embeddedness and the impact on performance.* Journal of International Business Studies, 2004: p. 428-442.
25. Steensma, H.K. and M.A. Lyles, *Explaining IJV survival in a transitional economy through social exchange and knowledge based perspectives.* Strategic Management Journal, 2000. **21**(8): p. 831-851.
26. Ho, R., *Handbook of univariate and multivariate data analysis and interpretation with SPSS*. 2006: CRC Press.
27. Goldberger, A.S., *Structural equation models: An overview.* Structural equation models in the social sciences, 1973: p. 1–18.
28. Pedhazur, E.J., *Multiple regression in behavioral research: Explanation and prediction.* 1997.
29. Chin, W.W., *Issues and opinion on structural equation modeling.* Management Information Systems Quarterly, 1998. **22**(1): p. 3.
30. Garson, G.D., *Statnotes: Topics in multivariate analysis.* Retrieved 10/05/2007 from http://faculty.chass.ncsu.edu/garson/PA765/statnote.htm, 2007.
31. Kline, T.J.B. and J.D. Klammer, *Path model analyzed with ordinary least squares multiple regression versus LISREL.* The Journal of Psychology: Interdisciplinary and Applied, 2001. **135**(2): p. 213-225.
32. Tabachnick, B.G. and L. Fidell, *Using multivariate statistics. ( 3rd. Ed).* New York: Harpercollins College Publishers, 1996.
33. Anderson, J.C. and D.W. Gerbing, *Structural equation modeling in practice: A review and recommended two-step approach.* Psychological bulletin, 1988. **103**(3): p. 411-423.
34. McFatter, R.M., *The use of structural equation models in interpreting regression equations including suppressor and enhancer variables.* Applied Psychological Measurement, 1979. **3**(1): p. 123.
35. Hair, J.F., et al., *Multivariate data analysis*. 1998: Prentice hall Upper Saddle River, NJ.
36. ISIk, Z., *A CONCEPTUAL PERFORMANCE MEASUREMENT FRAMEWORK FOR CONSTRUCTION INDUSTRY*, 2009, MIDDLE EAST TECHNICAL UNIVERSITY.